\documentclass[
amsmath,
amssymb,
amsfonts,
aps,
nofootinbib,
preprintnumbers,
prl,
reprint,
superscriptaddress,
a4paper,
draft
]{revtex4-2}
\usepackage[final]{graphicx}
\usepackage{xcolor}
\usepackage[
pdfauthor={Leron Borsten, Dimitri Kanakaris, Hyungrok Kim},
pdftitle={Three‐dimensional SL(2,ℝ) Yang–Mills theory is three‐dimensional gravity with background sources},
pdfsubject={hep-th,gr-qc},
breaklinks=true,
final
]{hyperref}
\usepackage[T1]{fontenc}
\usepackage{amsmath,amssymb,amsthm,booktabs,cleveref,orcidlink}
\usepackage[russian,english]{babel}
\usepackage{lmodern}
\usepackage[final,spacing]{microtype}

\begin{document}
\title{Three-dimensional \(\boldsymbol{\operatorname{SL}(2, \mathbb{R})}\)  Yang--Mills theory is three-dimensional gravity with background sources}
    \author{\firstname{Leron} \surname{Borsten}\orcidlink{0000-0001-9008-7725}}
    \email[]{l.borsten@herts.ac.uk}
    \affiliation{Department of Physics, Astronomy and Mathematics, University of Hertfordshire, Hatfield, Hertfordshire AL10 9AB, United Kingdom}
    \author{\firstname{Dimitri} \surname{Kanakaris}\orcidlink{0009-0001-7716-851X}}
    \email[]{d.kanakaris-decavel@herts.ac.uk}
    \affiliation{Department of Physics, Astronomy and Mathematics, University of Hertfordshire, Hatfield, Hertfordshire AL10 9AB, United Kingdom}
    \affiliation{Theoretische Natuurkunde, Vrije Universiteit Brussel, Pleinlaan 2, B-1050 Brussels, Belgium}
    \author{\firstname{Hyungrok} \surname{Kim}\orcidlink{0000-0001-7909-4510}}
    \email[]{h.kim2@herts.ac.uk}
    \affiliation{Department of Physics, Astronomy and Mathematics, University of Hertfordshire, Hatfield, Hertfordshire AL10 9AB, United Kingdom}
\begin{abstract}
Chern--Simons theory with certain gauge groups is known to be equivalent to a first-order formulation of three-dimensional Einstein gravity with a cosmological constant, where both  are purely topological. Here, 
we extend this correspondence to theories with dynamical degrees of freedom.
As an example, we show that three-dimensional Yang--Mills theory with gauge group \(\operatorname{SL}(2, \mathbb{R})\) is equivalent to the first-order formulation of three-dimensional Einstein gravity with no cosmological constant coupled to a background stress--energy tensor density (which breaks the diffeomorphism symmetry). The local degree of freedom of three-dimensional Yang--Mills theory corresponds to degenerate ``gravitational waves'' in which the metric is degenerate and the spin connection is no longer completely determined by the metric. Turning on a cosmological constant produces the third-way  (for \(\Lambda<0\)) or the imaginary third-way  (for \(\Lambda>0\)) gauge theories  with a background stress--energy tensor density.
\end{abstract}
\maketitle

\section{Introduction and summary}
Gauge theory and gravity are closely entwined by holographic duality \cite{Maldacena:1997re,Gubser:1998bc,Witten:1998qj} (reviewed in \cite{Aharony:1999ti,DHoker:2002nbb,Natsuume:2014sfa,Nastase:2015wjb}) and the double copy \cite{Bern:2008qj,Bern:2010ue,Bern:2010yg} (reviewed in~\cite{Carrasco:2015iwa,Borsten:2020bgv, Bern:2019prr,Adamo:2022dcm,Bern:2022wqg}). On the other hand, they differ in many ways: amount of symmetry (diffeomorphism versus Lie-algebra-valued gauge transformations), degrees of freedom (\(d-2\) versus \((d-3)d/2\) in \(d\) dimensions), etc.
In three dimensions, remarkably, pure gravity (with or without a  cosmological constant) is topological and known to be equivalent (at least perturbatively) to a topological gauge theory called Chern--Simons theory \cite{Achucarro:1986uwr,Witten:1988hc} (for reviews, see \cite{Carlip:1995zj,Carlip:1998uc,Banados:1998sm,Carlip:2005zn,Schroers:2007ey,Kiran:2014dfa,Carlip:2023nwa}); in this case, neither side has local degrees of freedom. 

This Letter shows that the above correspondence extends surprisingly to \emph{dynamical} theories. On the gauge theory side, we have ordinary Yang--Mills theory and its deformations,  the \emph{third-way theory} \cite{Arvanitakis:2015oga} (reviewed in \cite{Deger:2021ojb}), which occurs as a subsector of the ABJM theory on a Romans background, and its imaginary variant, which are all instances of the family of Manin theories \cite{Arvanitakis:2024dbu}. On the gravity side, we have three-dimensional Einstein gravity with a cosmological constant and a background stress--energy tensor density, much as one may turn on a classical background source density in Yang--Mills theory.
The background stress--energy tensor density breaks diffeomorphism invariance (just as a background Yang--Mills source breaks gauge invariance); if one allows degenerate metrics, this causes the spin connection to propagate with three degrees of freedom, matching those of  Yang--Mills theory with a gauge group of dimension three. Features of gauge theory correspond to features of gravity according to \cref{table:correspondence}.

Our construction does not involve matter (apart from the background stress--energy tensor density); note that  gauge-theoretic and gravitational minimal couplings to matter would be very different.
Furthermore, our construction is purely classical and, from the gravitational perspective, allows the dreibein and spin connection to be degenerate. Note, it has been argued that one must require the dreibein to be always invertible \cite{Witten:2007kt} and that this invertibility distinguishes Chern--Simons theory from three-dimensional gravity nonperturbatively. Moreover, our construction is perturbative in that we ignore large gauge transformations and the sum over topologies.

We work with Lorentzian-signature gravity, which has the effect that the Chern--Simons (and hence Yang--Mills) gauge groups do not admit positive-definite Killing forms and hence are not unitary. For unitarity, we may work with the gauge algebra \(\mathfrak{so}(4)=\mathfrak{su}(2)\oplus\mathfrak{su}(2)\), which corresponds formally to Euclidean gravity with a positive cosmological constant \cite{Witten:1988hc,Schroers:2007ey}. Note that the signature of the dynamical metric need not be the same as the signature of the background metric, so that \(\mathfrak {so}(4)\) Yang--Mills theory on a Lorentzian-signature background metric \(\hat g\) formally corresponds to Euclidean gravity with a Lorentzian-signature background stress--energy tensor density.

The gauge--gravity correspondence described here raises various possibilities, such as  analog gravity \cite{Visser:2001fe,Barcelo:2005fc} applications: condensed-matter realizations of Yang--Mills theory and other Manin theories may provide a laboratory for features of gravity involving propagating degrees of freedom. 

It may also have holographic implications. Chern--Simons theories on three-dimensional anti-de~Sitter space (AdS\textsubscript3) exhibit an AdS\textsubscript3/CFT\textsubscript2 duality with Wess--Zumino--Witten models \cite{Witten:1988hf,Coussaert:1995zp} (see reviews in \cite{Gawedzki:1999bq,Kraus:2006wn}). In particular, the asymptotic symmetries of AdS\textsubscript3 realize the Virasoro symmetry of the Wess--Zumino--Witten model, and the Bekenstein--Hawking entropy of the Ba\~nados--Teitelboim--Zanelli (BTZ) black hole corresponds to the Cardy entropy of two-dimensional conformal field theories. It is tempting to speculate that aspects of holography may extend to Yang--Mills theory seen as a gravitational theory. In the AdS\textsubscript3 case (third-way theory), the global \(\operatorname{Spin}(2,2)=\operatorname{SL}(2,\mathbb R)\times\operatorname{SL}(2,\mathbb R)\) isometry of AdS\textsubscript3 is broken by the masslike Manin term into the diagonal subgroup \(\operatorname{SL}(2,\mathbb R)_\mathrm{diag}\). This suggests that a holographically dual two-dimensional theory to the third-way theory, if one exists, should be such that instead of having left and right Virasoro symmetries, it should have only one (diagonal) copy of the Virasoro symmetry. If three-dimensional AdS Einstein gravity is dual to a monster CFT (as conjectured in \cite{Witten:2007kt}), then it may represent a deformation of that theory.

We use the \(-++\) metric signature with the Levi-Civita symbol \(\epsilon_{012}=-\epsilon^{012}\). Antisymmetrizations \(\dotsb_{[\mu}\dotsb_{\nu]}\) are always normalized.

\begin{table*}
\begin{center}
\begin{tabular}{ll} \toprule
3D gauge & 3D gravity \\\midrule
Chern--Simons & Einstein gravity + \(\Lambda\) \\
Yang--Mills & Einstein gravity + background \\
third way & Einstein gravity + \(\Lambda>0\) + background \\
imaginary third way & Einstein gravity + \(\Lambda<0\) + background \\
auxiliary field \(\tilde A^a_\mu\) & dreibein \(e^a_\mu\) \\
gauge field \(A^a_\mu\) & spin connection \(\frac12\epsilon^{abc}\omega_{bc\mu}\)\\
coupling constants \(g_\mathrm{YM},\lambda\) & coupling constants \(M_\mathrm{Pl},\Lambda\) \\
background densitized inverse metric \(\sqrt{|\det\hat g}(\hat g^{-1})^{\mu\nu}\) & background stress-energy tensor density \(T^{\mu\nu}\)\\
gauge symmetry partly broken by masslike term & diffeomorphism symmetry broken by background \\
unbroken gauge symmetry & rotation of dreibein index \\
vacuum & degenerate metric \\
Coulomb monopole & degenerate metric with singularity \\
pure gauge configuration & Minkowski vacuum \\
constant field strength configuration & dS/AdS vacuum \\
\bottomrule
\end{tabular}
\end{center}
\caption{Correspondence between gauge-theoretic concepts and gravitational concepts}\label{table:correspondence}
\end{table*}

\begin{table*}
\begin{tabular}{lll}\toprule
Max.\ symmetric vacuum & Manin pair & Manin theory \\\midrule
de~Sitter & \((\mathfrak{sl}(2,\mathbb C),\mathfrak{sl}(2,\mathbb R))\) & imaginary third-way \\
Minkowski & \((\mathrm T^*\mathfrak{sl}(2,\mathbb R),\mathfrak{sl}(2,\mathbb R))\) & 3d Yang--Mills \\
anti-de~Sitter & \((\mathfrak{sl}(2,\mathbb R)\oplus\mathfrak{sl}(2,\mathbb R),\mathfrak{sl}(2,\mathbb R)_\mathrm{diag})\) & third-way\\
\bottomrule 
\end{tabular}
\caption{Correspondence between Manin pairs, gauge theories, and sign of the cosmological constant}\label{table:manin-pairs}
\end{table*}

\section{Manin theories as gauge theories}
A \emph{Manin pair} \cite{Dri89} \((\mathfrak d,\mathfrak g)\) is a Lie algebra \(\mathfrak d\) equipped with an invariant nondegenerate inner product \(\langle-,-\rangle\) and a Lie subalgebra \(\mathfrak g\subset\mathfrak d\) whose dimension is half that of \(\mathfrak d\) and such that \(\langle x,y\rangle=0\) whenever \(x,y\in\mathfrak g\). The notion originates in the theory of integrable systems (see \cite{10.1007/BFb0113695} for a review).

A \emph{Manin theory} \cite{Arvanitakis:2024dbu} is a three-dimensional field theory associated to a Manin pair \((\mathfrak d,\mathfrak g)\) that is a deformation of a Chern--Simons theory valued in \(\mathfrak d\) with a masslike term:\footnote{This term, however, does not correspond to the masses of the propagating degrees of freedom, which are massless.}
\begin{equation}
	S[\mathbb A] = \int\tfrac12k\left\langle\mathbb A,\mathrm d\mathbb A + \tfrac13[\mathbb A\wedge\mathbb A]\right\rangle+\tfrac12\left\langle\mathbb A,\hat\star M\mathbb A\right\rangle
\end{equation}
where \(\mathbb A\) is a \(\mathfrak d\)-valued connection and \(M\colon \mathfrak d\to\mathfrak d\)
is a linear map of mass dimension \(1\) with kernel and image both equal to \(\mathfrak g\subset\mathfrak d\) such that
\begin{align}
    \langle My,z\rangle&=\langle y,Mz\rangle&
    M[x,y]&=[x,My]
\end{align}
for \(x\in\mathfrak g\) and \(y,z\in\mathfrak d\). The Hodge star \(\hat\star\) is taken with respect to a background (pseudo-)Riemannian metric \(\hat g\). The masslike term \(\frac12\langle\mathbb A\wedge\hat \star M\mathbb A\rangle\) breaks diffeomorphism symmetry (since it involves the background metric \(\hat g\)) and the part of the gauge symmetry \(\mathfrak d\) not contained in \(\mathfrak g\).\footnote{This does not affect the consistency of the theory since half of the fields are rendered auxiliary so that the associated gauge symmetry is no longer necessary.} The background Hodge star structure coefficients are given by the (1,2)-tensor \(\sqrt{|\det\hat g|} \hat g^{\mu\sigma} \epsilon_{\sigma\nu\rho}\). We may regard it as the densitized symmetric (2,0)-tensor $ \sqrt{|\det\hat g|}\hat g^{\mu\sigma}$, similar to a stress--energy tensor density.

We may choose a linear subspace \(\tilde{\mathfrak g}\) inside \(\mathfrak d\) such that \(\mathfrak d=\mathfrak g+\tilde{\mathfrak g}\) and \(\langle-,-\rangle\) identifies \(\tilde{\mathfrak g}\cong\mathfrak g^*\).\footnote{This choice does not affect the physics \cite{Arvanitakis:2024dbu}.} Picking a basis \(\{t_a\}\) of \(\mathfrak g\) and a corresponding basis \(\{\tilde t^a\}\) of \(\tilde{\mathfrak g}\) such that $\langle\tilde t^a,t_b\rangle = \delta^a_b$, we may parameterize the structure constants of the Manin pair and the operator \(M\) as
\begin{equation}
\begin{aligned}
    [t_a,t_b]&=f_{ab}{}^ct_c&
    [\tilde t^a,\tilde t^b]&=\tilde f^{ab}{}_{c}\tilde t^c+\tilde h^{abc}t_c\\
    [t_a,\tilde t^b]&=\tilde f^{bc}{}_at_c-f_{ac}{}^b\tilde t^c &
    M\tilde t^a&=M^{ab}t_b.
\end{aligned}
\end{equation}
Then the \(\mathfrak d\)-valued gauge field \(\mathbb A\) decomposes as \(\mathbb A=A^at_a+\tilde A_a\tilde t^a\), its respective $\mathfrak g$- and $\tilde{\mathfrak g}$-valued components, in terms of which the action becomes
\begin{multline}
    S[A,\tilde A] =
    \int k\tilde A_a\wedge F^a + \tfrac12k\tilde f^{ab}{}_c\tilde A_a\wedge\tilde A_b\wedge A^c \\+ \tfrac16k\tilde h^{abc}\tilde A_a \wedge\tilde A_b \wedge\tilde A_c +\tfrac12M^{ab}\tilde A_a\wedge\hat\star\tilde A_b,
\end{multline}
where we have defined the field strength \(F = F^at_a = \mathrm dA + A\wedge A\) whose components are \(F^a=\mathrm dA^a+\frac12f_{bc}{}^aA^b\wedge A^c\).

We now specialize to the case of a gravity-like Manin theory, where we now take the Manin pair to be one of
\begin{equation}\label{eq:gravitational_manin_pairs}
	(\mathfrak d,\mathfrak g) =
	\begin{cases}(\mathfrak{sl}(2,\mathbb R)\oplus\mathfrak{sl}(2,\mathbb R),\mathfrak{sl}(2,\mathbb R)_\mathrm{diag}) \\ (\mathrm T^*\mathfrak{sl}(2,\mathbb R),\mathfrak{sl}(2,\mathbb R)) \\ (\mathfrak{sl}(2,\mathbb C),\mathfrak{sl}(2,\mathbb R)), \end{cases}
\end{equation}
where \(\mathrm T^*\mathfrak{sl}(2,\mathbb R)\) means the semidirect sum \(\mathfrak{sl}(2,\mathbb R)\ltimes\mathfrak{sl}(2,\mathbb R)^*\) with the dual space \(\mathfrak{sl}(2,\mathbb R)^*\) regarded as an Abelian Lie algebra and acted upon by \(\mathfrak{sl}(2,\mathbb R)\) as the coadjoint representation.
These three cases correspond respectively to the third-way theory, Yang--Mills theory, and the imaginary third-way theory according to \cref{table:manin-pairs}.
Normalizing the \(\mathfrak g\cong\mathfrak{sl}(2,\mathbb R)\) generators as
\begin{equation}
	[t_a,t_b] = \epsilon_{abc}\eta^{cd}t_d = \epsilon_{abc}t^c,
\end{equation}
where \(a,b,c,d\in\{0,1,2\}\) are raised and lowered by \(\eta_{ab}=\operatorname{diag}(-1,1,1)\), then \(\mathfrak d\) and \(M\) have the corresponding structure constants
\begin{equation}
\begin{aligned}
	[t_a,t_b] &= \epsilon_{abc}\eta^{cd}t_d
	&
	[t_a,\tilde t^b] &= \eta^{bc}\epsilon_{acd}\tilde t^d \\
	[\tilde t^a,\tilde t^b] &= \tfrac14\nu\epsilon^{abc}t_c
 	&
  	M\tilde t^a&=\mu\eta^{ab}t_b,
\end{aligned}
\end{equation}
where \(\mu\) is a constant of mass dimension 1 and
\begin{equation}\label{eq:nu}
    \nu=\begin{cases}
        1 & \text{if \(\mathfrak d=\mathfrak{sl}(2,\mathbb R)\oplus\mathfrak{sl}(2,\mathbb R)\)} \\
        0 & \text{if \(\mathfrak d=\mathrm T^*\mathfrak{sl}(2,\mathbb R)\)} \\
	-1 & \text{if \(\mathfrak d=\mathfrak{sl}(2,\mathbb C)\)}.
    \end{cases}
\end{equation}
Accordingly, their respective Manin gauge theory actions are
\begin{multline}\label{eq:gauge-grav-action}
	S[A,\tilde A] = \int(k\tilde A_a\wedge F^a + \tfrac1{24}\nu k\epsilon^{abc}\tilde A_a\wedge \tilde A_b\wedge \tilde A_c \\ + \tfrac12\mu\eta^{ab}\tilde A_a\wedge\hat\star\tilde A_b).
\end{multline}
The field \(\tilde A\) is auxiliary in that its equation of motion
\begin{equation}
    \tilde A_a = k\mu^{-1}\eta_{ab}\hat\star\left(F^b +\tfrac18\nu\epsilon^{bcd}\tilde A_c\wedge\tilde A_d\right)
\end{equation}
can be recursively substituted as
\begin{equation}
	\tilde A_a
	=
	\frac k\mu\eta_{ab}\hat\star F^b + \frac{k^3}{8\mu^3}\nu\epsilon_{abc}\hat\star F^b\wedge\hat\star F^c 
	+ \mathcal{O}((k/\mu)^4\nu^2F^3) 
\end{equation}
to produce an action
\begin{equation}
    S[A] = \int \tfrac12k^2\mu^{-1}\eta_{ab}F^a\wedge\hat\star F^b + \mathcal{O}(k^4\mu^{-3}F^3).
\end{equation}
If we define
\begin{align}
    g_\mathrm{YM}&=\sqrt{|\mu|}/k&
    \lambda&=k^4|\mu|^{-3/2},
\end{align}
then we get
\begin{equation}\label{eq:manin-gauge}
    S[A] = \int \operatorname{sgn}(\mu)\frac1{2g_\mathrm{YM}^2}\eta_{ab}F^a\wedge\hat\star F^b + \mathcal{O}(\nu\lambda g_\mathrm{YM}^{-3}F^3),
\end{equation}
realizing it as a nonlinear correction to Yang--Mills theory.\footnote{The theory is not unitary for either sign of \(\mu\) since the Killing form for \(\mathfrak{sl}(2,\mathbb R)\) is indefinite.}
In particular, when \(\nu=0\) or when \(g_\mathrm{YM}\) is fixed and \(\lambda\to0\), we obtain pure Yang--Mills theory. The corrections are interesting in that they give rise to local field theories, but this locality is not manifest without the use of auxiliary fields (such as \(\tilde A\)); while one can write down the equations of motion in a manifestly local form \citep{Arvanitakis:2015oga}, the source current that describes the higher-order self-interactions is only conserved on shell. Such theories are referred to as ``third-way consistent'' \citep{Bergshoeff:2014pca,Arvanitakis:2014yja,Arvanitakis:2014xna}.

\section{Manin theories as gravity theories}
The Manin theories associated to the Manin pairs \eqref{eq:gravitational_manin_pairs}, which describe dynamical gauge fields atop a fixed background metric, admit a reinterpretation as gravitational theories describing a dynamical metric coupled to a background matter gauge field. 

To make the gravitational interpretation manifest, we let \(M_\mathrm{Pl}\) be an arbitrary mass scale and make the following identifications. Define the dreibein $e^a$, the metric \(g_{\mu\nu}\), the spin connection $\omega^{ab}$, the cosmological constant \(\Lambda\), and the stress--energy tensor density \(T^{\mu\nu}\) as
\begin{subequations}
\begin{align}
\label{dreibein spin connection}
	e^a &= k\tilde A^a/M_\mathrm{Pl}\\
	g_{\mu\nu} &= \eta_{ab}e^a_\mu e^b_\nu\\
	\omega^{ab} &= -\epsilon^{abc}A_c
	\iff
	A^a = \tfrac12\epsilon^{abc}\omega_{bc}\\\label{eq:nulambda}
	\Lambda &= -\tfrac14k^{-2}\nu M_\mathrm{Pl}^2\\
	T^{\mu\nu} &= k^{-2}\mu M_\mathrm{Pl}^2(\hat g^{-1})^{\mu\nu}\sqrt{|\det\hat g|}.
\end{align}
\end{subequations}
Note that \(T^{\mu\nu}\) is a tensor density rather than a tensor; it should be thought of as the stress--energy tensor density of the background matter.\footnote{In many ways, the stress--energy tensor density is more fundamental than the stress--energy tensor --- e.g.\ the definition \(-2\delta S_\mathrm{matter}/\delta g^{\mu\nu}\) is naturally a density.}
With these definitions, the Manin theory action \eqref{eq:gauge-grav-action} now takes the gravitational form
\begin{multline}
	S[e,\omega] = \tfrac12M_\mathrm{Pl} \int \epsilon_{abc}\left(e^a\wedge R^{bc}-\tfrac13\Lambda e^a\wedge e^b\wedge e^c\right) \\\qquad+\tfrac12\int\mathrm d^3x\, T^{\mu\nu}\eta_{ab}e^a_\mu e^b_\nu,
\end{multline}
where the Riemann curvature is expressed as the 2-form
\begin{equation}
	R^{ab} = \mathrm d\omega^{ab} + \omega^a{_c}\wedge\omega^{cb}.
\end{equation}

The three choices of Manin pair \eqref{eq:gravitational_manin_pairs}, encoded by $\nu=1,0,-1$ in \eqref{eq:nu}, now correspond via \eqref{eq:nulambda} to  $\Lambda<0, \Lambda=0, \Lambda>0$, respectively. In the original gauge theory picture given in \eqref{eq:gauge-grav-action} and \eqref{eq:manin-gauge}, the (anti-)de Sitter cases correspond to the third-way theories of \citep{Bergshoeff:2014pca,Arvanitakis:2014yja,Arvanitakis:2014xna}. When the cosmological constant is vanishing, $\nu=0=\Lambda$,  the Manin theory \eqref{eq:gauge-grav-action} reduces to the familiar first-order formulation of pure Yang--Mills theory with $\operatorname{SL}(2, \mathbb{R})$ gauge group and,  correspondingly, the higher-order terms in \eqref{eq:manin-gauge} are switched off.

The equations of motion are
\begin{subequations}
\begin{align}
	\mathrm de^a + \omega^a{}_b\wedge e^b &= 0 \\\label{eq:torsionfree}
	R^{ab}{}_{\mu\nu}-2\Lambda e^a_{[\mu}e^b_{\nu]} - 
	M_\mathrm{Pl}^{-1}\epsilon_{\mu\nu\rho}T^{\rho\sigma}e^c_\sigma\epsilon^{ab}{}_c &= 0,
\end{align}
\end{subequations}
where the first implies  the spin connection be torsion-free and the second is Einstein's field equations with a cosmological constant and a background stress--energy term.

Suppose that \(e_\mu^a\) is invertible.\footnote{This is not an innocuous assumption: for example, a Coulomb monopole in some Abelian subalgebra \(\mathfrak u(1)\subset\mathfrak{sl}(2,\mathbb R)\) will not be such that the dreibein is invertible.}
We now raise and lower Lorentz indices via \(g_{\mu\nu}\).
The torsion-free condition \eqref{eq:torsionfree}  can be solved to yield the usual definition for the spin connection
\begin{equation}\label{eq:dreibein-to-spin-connection}
    \omega_\mu^{ab}=e^{\nu[a}(\partial_\mu e_\nu^{b]}-\partial_\nu e_\mu^{b]}+e^{\sigma |b]}e_\mu^c\partial_\sigma e_{\nu c}).
\end{equation}

As to the second, let us (again using invertibility of the dreibein) convert all dreibein indices to ordinary vector indices:
\begin{equation}
    R_{\mu\nu\rho\sigma} - \Lambda(g_{\mu\rho}g_{\nu\sigma}-g_{\nu\rho}g_{\mu\sigma}) - 
    \frac{\sqrt{|\det g|}}{M_\mathrm{Pl}}\epsilon_{\mu\nu\tau}\epsilon_{\rho\sigma\lambda}T^{\tau\lambda} = 0.
\end{equation}
All terms now carry four Lorentz indices, which have the symmetry of the Young tableau \(\boxplus\). In three dimensions, we lose no information  contracting the  \(\nu\) and \(\sigma\) indices\footnote{This is the familiar fact that, in three dimensions, the Riemann tensor is algebraically determined by the Ricci tensor alone.} to obtain
\begin{equation}
    0=R_{\mu\nu}-2\Lambda g_{\mu\nu}
    -\frac{M_\mathrm{Pl}^{-1}}{\sqrt{|\det g|}}(T_{\mu\nu}-T_{\rho\sigma}g^{\rho\sigma}g_{\mu\nu}).
\end{equation}
We can equivalently rewrite this as
\begin{equation}\label{eq:einstein_field_equation}
    \sqrt{|\det g|}(G_{\mu\nu}+\Lambda g_{\mu\nu})=M_\mathrm{Pl}^{-1}T_{\mu\nu},
\end{equation}
where
\begin{equation}
    G_{\mu\nu} = R_{\mu\nu}-\tfrac12g_{\mu\nu}R
\end{equation}
is the usual Einstein tensor.

\section{Vacuum solutions}
From the gauge-theory perspective, the gauge-theoretic vacuum configuration is simply one in which all fields vanish:
\begin{align}
A^a_\mu(x)&=0=\tilde A^a_\mu(x).
\end{align}
From a gravitational perspective, however, this corresponds to a degenenerate metric \(g_{\mu\nu}=0=\omega^{ab}_\mu\).

On the other hand, there exist a family of solutions that naturally correspond to the notion of a gravitational vacuum.
Suppose that the background metric \(\hat g_{\mu\nu}=\hat e_\mu^a\hat e_{a\nu}\) satisfies Einstein's equations with cosmological constant \(\Lambda^*\) --- for example, it can be Minkowski, dS or AdS. Take the dynamical dreibein to be proportional to the background dreibein:
\begin{equation}
	e^a_\mu = \alpha\hat e^a_\mu.
\end{equation}
Then it is straightforward to check that this is a solution to the Manin theory provided that
\begin{equation}\Lambda^*=\alpha^2\Lambda+\alpha M_\mathrm{Pl}\mu.\end{equation}
In this manner all standard solutions are recovered but with a shifted cosmological constant.  

\section{Gauge/gravitational waves}
The reader may wonder what happened to the local degrees of freedom: three-dimensional Yang--Mills theory with gauge group \(\mathfrak g\) has \(\dim(\mathfrak g)\) local degrees of freedom, while three-dimensional pure Einstein gravity should have no local degrees of freedom. In fact, the latter statement depends on whether one is working with the first-order or second-order formulations of Einstein gravity.

Seen as a gauge theory, Yang--Mills theory on a Minkowski background \(\hat g_{\mu\nu}=\eta_{\mu\nu}\) has the plane-wave solution
\begin{subequations}
\begin{align}
    A^a_\mu(x)&=\varepsilon_\mu C^a\exp(\mathrm ik\cdot x)\\
    \tilde A^a_\mu(x)&=
    k\mu^{-1}\epsilon_\mu{}^{\nu\rho}\mathrm ik_{[\nu}\varepsilon_{\rho]}C^a\exp(\mathrm ik\cdot x)
\end{align}
\end{subequations}
for a fixed polarization vector \(\varepsilon_\mu\) such that \(k\cdot\varepsilon=0\) (with two polarization vectors \(\varepsilon,\varepsilon'\) gauge-equivalent if \(\varepsilon_\mu-\varepsilon'_\mu\propto k_\mu\)) and colour \(C\in\mathfrak{sl}(2,\mathbb R)\). Translated into gravitational language, we obtain
\begin{subequations}
\begin{align}
    \omega^{ab}_\mu&=-\epsilon^{abc}C_c\varepsilon_\mu\exp(\mathrm ik\cdot x)\\
    e^a_\mu&=k^2M_\mathrm{Pl}^{-1}\mu^{-1}\epsilon_\mu{}^{\nu\rho}\mathrm ik_{[\nu}\varepsilon_{\rho]}C^a\exp(\mathrm ik\cdot x).
\end{align}
\end{subequations}
That is, the plane waves correspond to degenerate solutions in which the dynamical metric is noninvertible and the spin connection remains nonzero. This is possible in the first-order formalism since the relation \eqref{eq:dreibein-to-spin-connection} between the spin connection and the dreibein assumes that the dreibein is invertible; when this fails, the spin connection cannot be solved in terms of the dreibein.
Thus, first-order Einstein gravity acquires three degrees of freedom in the presence of a background stress--energy tensor density if one allows degenerate metrics.

\section{Axisymmetric solutions: Coulomb monopoles and black holes}
From the gauge-theoretic point of view, we have solutions that correspond to  spherically symmetric Coulomb monopoles:
\begin{align}
    e^a(r) &= Q^a \ln r &
    \omega^{ab}(r) &= \mu^{-1}\epsilon^{ab}{}_cQ^c r^{-1}\star\mathrm dr
\end{align}
where \(Q\in\mathfrak{sl}(2,\mathbb R)\) is a fixed direction and \(r\) is the spatial radial coordinate in a cylindrical coordinate system.

From a gravitational point of view, these correspond to the degenerate dynamical metrics
\begin{equation}
    g_{\mu\nu}(t,r,\theta)=\begin{pmatrix}
        0 & 0 & 0\\
        0 & Q^2 \ln r & 0 \\
        0 & 0 & 0
    \end{pmatrix}.
\end{equation}
These  seem very pathological from the gravitational perspective, but  are perfectly well-behaved elementary solutions from the Yang--Mills point of view.

When solved with an axisymmetric ansatz, the Einstein field equations for three-dimensional Einstein gravity with a negative cosmological constant admit the BTZ black hole solution \cite{Banados:1992wn} (reviewed in \cite{Brill:1998pr,Carlip:1995qv,Carlip:1998qw,Banados:1998gg}).
Let us examine a static axisymmetric ansatz for our case with a background stress--energy tensor density (assumed to be also static and axisymmetric, e.g.~for a Minkowski background).
Assuming invertibility of the dreibein, we may postulate the ansatz
\begin{equation}
    g_{\mu\nu}\mathrm dx^\mu\mathrm dx^\nu=u(r)^2\,\mathrm dt^2+v(r)^{-2}\,\mathrm dr^2+w(r)^2\,\mathrm d\theta^2
\end{equation}
for the dynamical metric, whose Einstein tensor density is then
\begin{subequations}
\begin{align}
    \sqrt{|\det g|}G^{tt}&=-u^{-1}(vw')'\\
    \sqrt{|\det g|}G^{rr}&=v^3u'w'\\
    \sqrt{|\det g|}G^{\theta\theta}&=w^{-1}(u'v)'.
\end{align}
\end{subequations}
(Here \('\) denotes \(\mathrm d/\mathrm dr\).)
For a flat background stress--energy tensor density
\begin{equation}
    T^{\mu\nu}=\operatorname{diag}(-a(r),b(r)^{-1},c(r))
\end{equation}
and cosmological constant \(\Lambda\), we have the equations
\begin{align}
-u^{-1}(vw')'-\Lambda u^{-1}w/v&=-a\\
v^3u'w'+\Lambda uvw&=b^{-1}\\
w^{-1}(u'v)'+\Lambda uw^{-1}/v&=c.
\end{align}
Defining \(f=vw'\) and \(g=u'v\), we may solve for \(v\) in terms of \(u,w,f,g\) as
\begin{equation}
    v=(b(fg+\Lambda uw))^{-1},
\end{equation}
so that we obtain the system of first-order ordinary differential equations
\begin{subequations}\label{eq:numerical-system}
\begin{align}
f'&=au-\Lambda bw(fg+\Lambda uw)\\
g'&=cw-\Lambda bu(fg+\Lambda uw)\\
u'&=bg(fg+\Lambda uw)\\
w'&=bf(fg+\Lambda uw).
\end{align}
\end{subequations}
This is a Hamiltonian system with generalized coordinates \((f,g)\), generalized momenta \((u,w)\), and Hamiltonian
\begin{equation}
    H=\tfrac12au^2+\tfrac12cw^2-\tfrac12b(fg)^2-\Lambda bfguw-\tfrac12\Lambda^2b(uw)^2.
\end{equation}
We can often reparameterize the radial coordinate so as to have \(b=1\), in which case the Hamiltonian simplifies to
\begin{equation}
    H=\tfrac12au^2+\tfrac12cw^2-\tfrac12(fg+\Lambda uw)^2.
\end{equation}
This describes a nonrelativistic particle moving in two spatial directions with metric \(\operatorname{diag}(a^{-1},c^{-1})\) and a momentum-dependent potential \(V=-\frac12(fg+\Lambda uw)^2\), which is unbounded below. A black-hole-like apparent singularity occurs at \(V=-\infty\), which is at the top of the potential hill \(V\). Solutions therefore tend to roll down to \(v\to\infty\) to run into apparent singularities.\footnote{By \eqref{eq:einstein_field_equation}, these singularities are always apparent unless the background \(T^{\mu\nu}\) is itself singular.} On the other hand, to have the solution be nonsingular at spatial infinity for a general \(T^{\mu\nu}\), one may need to tune parameters so as to end up at the top of the hill.\footnote{Of course, one can always craft \(T^{\mu\nu}\) so as to obtain black hole solutions --- in particular, when the background metric \(\hat g\) is already a black-hole geometry.} Note that a naïve count of the initial conditions of the system \eqref{eq:numerical-system} indicates more parameters than would be expected from the no-hair theorem.

    \begin{acknowledgments}
        \noindent \emph{Acknowledgments.} H.K. was supported by the Leverhulme Research Project Grant RPG--2021--092. The authors thank Alexandros Spyridion Arvanitakis\orcidlink{0000-0002-7844-5574} for helpful comments. H.K. thanks David Simon Henrik Jonsson\orcidlink{0009-0001-7155-8496} for helpful comments.
    \end{acknowledgments}

\bibliographystyle{unsrturl}
\bibliography{biblio}
\end{document}